\documentclass[12pt]{article}

\usepackage[utf8]{inputenc} 
\usepackage{amsmath, amssymb} 
\usepackage{graphicx} 
\usepackage{hyperref} 
\usepackage{geometry} 
\usepackage{float}
\usepackage{caption}
\usepackage{subcaption}
\usepackage{comment}
\usepackage{xcolor}
\usepackage{authblk}

\title{Universal features of non-analytical energy storage in quantum critical quantum batteries}

\author[1,2]{Riccardo Grazi}
\author[1,2]{Dario Ferraro}
\author[1,2]{Niccolò Traverso Ziani}

\affil[1]{\small Dipartimento di Fisica, Università degli Studi di Genova, via Dodecaneso 33, 16146, Genova, Italy}
\affil[2]{\small CNR SPIN, via Dodecaneso 33, 16146, Genova, Italy}
\date{}
\begin{document}

\maketitle

\begin{abstract}
Quantum batteries are quantum mechanical systems able to store and release energy in a controlled fashion. Among them, a special role is played by quantum structures defined as networks of two-level systems. In this context, it has recently been shown that the energy stored in free fermion quantum batteries is sensitive to the quantum phase diagram of the battery itself. This sensitivity is relevant for stabilizing the stored energy and designing optimal charging protocols. In this article, we explore universal charging behaviors of free fermion quantum batteries across quantum phase transitions. We first analyze a Dirac cone-like model to extract general features. Then, we verify our findings by means of two relevant lattice models, namely the Ising chain in a transverse field and the Haldane model.
\end{abstract}

\section{Introduction}
As witnessed by the 2025 Nobel prize in physics, conceptual and technological advancements allow for macroscopic systems to show quantum behavior~\cite{Frowis18}. This fact offers an impressive range of possibilities, among which a special role is played by the development of quantum technologies~\cite{Aguado24}. In this context, the hope is that quantum systems will provide innovative solutions in a range of fields that spans from sensing to computing, through communication and simulation~\cite{Benenti_book}. To advance, the field of quantum technology needs quantum building blocks with different functionalities. Among such building blocks, quantum batteries are gaining considerable attention, both from the theoretical and the experimental point of view~\cite{Bhattacharjee21, Quach23, Campaioli24}. They are quantum systems able to store energy, and release it in the form of work, at request~\cite{Alicki13}. They are currently object of intense research since they are relevant under many perspectives. On one hand, it has been shown that they can show superextensive power charging, namely they can charge faster and faster as the number of constituents in the battery is increased~\cite{Binder15, Campaioli17, Ferraro18, Julia20, Rossini20, Gyhm22, Dou22, Dou22b}. On the other hand, it has been conjectured that having quantum energy sources can be useful in the control of quantum devices, thanks to a matching of energy and time scales~\cite{Chiribella21, Cioni24, Kurman25}. Moreover, the inspection of quantum batteries shares intruguing intersections with fundamental issues in quantum thermodynamics of both closed and open quantum systems~\cite{Vinjanampathy16, Deffner19, Carrega16, Farina19, Carrega22, Morrone23, Yang24, Elyasi25, Verma25b}, where it for instance lead to the definition of the concept of ergotropy~\cite{Allahverdyan04}. Such important figure of merit quantifies the amount of work that can be extracted from a quantum system by means of a unitary operations \cite{DiBello25, Formicola25, Zahia25, Hadipour25, Liu2025, Yao2025}.\\
At the practical level, quantum batteries can be classified into two large classes: batteries represented by systems with a small number of quantum degrees of freedom~\cite{Le18, Zhang19, Hu22, Gemme22, Catalano24, Chand25, Evangelakos25, Zhang25, Razzoli25, LiWu25, Hu25, Siddique25}, and batteries with a large number of quantum degrees of freedom~\cite{Seah21, Shaghaghi22, Quach22, Grazi24, Grazi25, Hymas25, Massa25}. While the first scenario allows for the (numerically) exact solution of complex problems, such as finding optimal charging regimes and Hamiltonian couplings~\cite{Rodriguez23,Erdman24,Gao25, Zahia2025optimizing, Sun24}, or for the consideration of the presence of dissipation to reservoirs~\cite{Barra19, Hovhannisyan20, Cavaliere25, Zhang25, Yao2025, ZhaoAug2025, Liu2025, Liu25}, the second merges the complexity of many body quantum physics with the ones of quantum thermodynamics and exact results are more rare~\cite{Franchini16}. Still, large quantum batteries are relevant both because they can store more energy and because they can show the already mentioned superextensive charging power.
In the context of large quantum batteries, many practical tools are provided by the theory of quantum quenches~\cite{Mitra18}, recently developed for- among other reasons- understanding quantum thermalization or the lack thereof. In such a theory, a remarkable role is played by the presence of quantum phase transitions~\cite{Sachdev11,dqpt1,dqpt2,wigner}. Indeed, quantum quenches across quantum phase transitions manifest intruguing phenomena such as dynamical quantum phase transitions~\cite{dqpt1,dqpt5,dqpt4,dqpt3,dqpt6,dqpt2,dqpt7,Porta20,dqpt8,dqpt9,dqpt10} and stady states with large deviations with respect to thermal states. It is hence natural to think that quantum phase transitions can play a role in the context of quantum battery charging as well~\cite{Ferraro19, Barra22, Grazi24}.\\
This is indeed the case. For instance, the following scenario has been recently analyzed~\cite{Grazi25}.\\
One considers a quantum battery with Hamiltonian $H_B$ given by a collection of free fermions with definite (quasi-)momentum, namely
\begin{equation}
H_B=\sum_{\mathbf{k}(\in BZ)} (c^\dagger_{a,\mathbf{k}},c^\dagger_{b,\mathbf{k}})(\mathbf{d}^A(\mathbf{k})\cdot \boldsymbol{\sigma})(c_{a,\mathbf{k}},c_{b,\mathbf{k}})^T. \label{H_Gen_Non_Superconduttiva}
\end{equation}
Here, $\mathbf{k}$ is the (quasi)-momentum, $BZ$ indicates the Brillouin zone, $\mathbf{d}^A(\mathbf{k})=({d}_1^A(\mathbf{k}),{d}_2^A(\mathbf{k}),{d}_3^A(\mathbf{k}))$ are free parameters, and $\boldsymbol{\sigma}$ is the Pauli matrix vector in the usual representation respectively. Finally $c_{\nu,\mathbf{k}}$, with \textcolor{black}{$\nu=a,b$}, is the fermionic annihilation operator for a fermion in the quantum state labelled by $\nu$ and by the (quasi)-momentum $k$. The battery is prepared, for time $t<0$, in the groud state of its Hamiltonian. At time $t=0$ a parameter in the Hamiltonian is suddenly varied to induce a charging process. Such a charging lasts for a time $\tau$, after which the system is governed again by the initial Hamiltonian $H_B$. In other words, the evolution of the system for $0<t<\tau$ is controlled by the Hamiltonian $H_E$ given by
\begin{equation}
H_E=\sum_{\mathbf{k}(\in BZ)} (c^\dagger_{a,\mathbf{k}},c^\dagger_{b,\mathbf{k}})(\mathbf{d}^B(\mathbf{k})\cdot \boldsymbol{\sigma})(c_{a,\mathbf{k}},c_{b,\mathbf{k}})^T,
\end{equation}
where $(\mathbf{d}^B(\mathbf{k})=({d}_1^B(\mathbf{k}),{d}_2^B(\mathbf{k}),{d}_3^B(\mathbf{k}))$ encodes the quench performed to achieve the charging. What is finally inspected is the energy $\Delta E(\tau)$ stored in the system for $t>\tau$. While we refer to Ref.~\cite{Grazi25} for the details of the derivation, it is useful to state here the functional form of such quantity. It is given by
\begin{equation}
\Delta E(\tau)=\sum_{\mathbf{k}\in BZ}\frac{1-\cos(2\omega(\mathbf{k})\tau)}{\omega(\mathbf{k})^2\epsilon(\mathbf{k})}F_0(\mathbf{k}), \label{eq:FormulaGenerale_NON_SC}
\end{equation}
where
\begin{eqnarray}
F_0(\mathbf{k})&=&\frac{\omega(\mathbf{k})^2}{\omega(\mathbf{k})^2-\left(d^B_3(\mathbf{k})\right)^2}\left(d^A_1(\mathbf{k})d^B_2(\mathbf{k})-d^A_2(\mathbf{k})d^B_1(\mathbf{k})\right)^2 \nonumber\\ &+& \bigg( d^A_3(\mathbf{k}) \sqrt{\omega(\mathbf{k})^2-\left( d^B_3(\mathbf{k})\right)^2}\nonumber\\ &-&\frac{d^B_3(\mathbf{k})}{\sqrt{\omega(\mathbf{k})^2-\left( d^B_3(\mathbf{k})\right)^2}}\left(d^A_1(\mathbf{k})d^B_1(\mathbf{k})+d^A_2(\mathbf{k})d^B_2(\mathbf{k})\right)\bigg)^2.\nonumber\\
\, \label{F0}
\end{eqnarray}
Here, $\epsilon(\mathbf{k})$ and $\omega(\mathbf{k})$ denote the dispersion relations of the Hamiltonian before the first quench and during the intermediate evolution, respectively. We consider $\tau$ large enough that the cosine term averages out and can be neglected. \textcolor{black}{The reason why this approximation is physically justified is that, as showed in \cite{Grazi24, Grazi25}, the early-time oscillations that affect the stored energy as function of $\tau$ decay due to the summation over $k$, allowing energy to reach a plateau which is stable with respect to temporal oscillations.} Clearly, when either $H_B$ or- more drastically- $H_E$ are critical, \textcolor{black}{i.e. when either Hamiltonian is tuned to a critical point associated with a quantum phase transition,} the stored energy runs in a 0/0 form that can result in non-analytical dependencies of the energy stored with respect to the charging parameters. This fact hence marks the strong dependence of the stored energy onto the quantum phase diagram~\cite{Grazi24}.\\
In this work, we characterize the universal features that quantum phase transitions imprint onto the energy stored in quantum batteries whose working mechanism is the one just described, and for which the quantum phase transition is described by a Dirac cone structure. We find that in one dimension the first derivative of the stored energy is discontinuous, while in two dimensions the second derivative diverges. We corroborate our finding by means of two relevant lattice models, namely the quantum Ising model in a transverse field~\cite{Pfeuty70}, and the two-dimensional Haldane model~\cite{Haldane88,h2,h3,h4,h5}. \textcolor{black}{We emphasize that our results applies to models whose Hamiltonians can be written in terms of non-interacting fermions and in which the quenched parameter drives the system through a linear Dirac-cone gap closing. Different scenarios, such as interacting systems or gap closings with non-linear dispersions, may exhibit different non-analytic scalings. However, whenever a system features a linear band touching and the quench pushes the post-quench Hamiltonian across that point, our analytic predictions hold universally, regardless of which microscopic parameter is actually quenched.}\\
The rest of the article is structured as Follows: In Sec.2. we describe the Dirac cone model and derive the relative results, In Sec.3 we compare the result obtained with the ones relative to the Ising and Haldane model. Finally, in Sec.4, we draw our conclusions.
\section{Dirac models} \label{Dirac}
The non analytical behavior of the energy stored into the quantum battery comes from the vicinity of the gap closing point in the dispersion relation. Consequently, to understand its universal features, we consider the paradigmatic low energy models for one- and two-dimensional non-interacting systems undergoing quantum phase transitions with a linear gap closing~\cite{Sachdev11}. We hence adopt, as quantum battery Hamiltonians, the Hamiltonians (respectively $d=1,2$ for one or two spatial dimensions)
\begin{equation}
    H_{d D} = \mathbf{d_d}\cdot \boldsymbol{\sigma} \label{GeneralHam}
\end{equation}
where $\mathbf{d_1}=(k_x,0, m_A)$ and $\mathbf{d_2}=(k_x, k_y, m_A)$. Here $k_x$ and $k_y$ label momentum in the $x$ and $y$ directions respectively, and $m_A$ is the mass. The charging is implemented by quenching the mass from $m_A$ to $m_B$. The system is critical when one of the two masses is zero. Moreover, since we deal with a low energy expansion, we replace the discrete sum with an integral in $k$ space- here restricted to a shell of radius $\Lambda$ around zero momentum. \textcolor{black}{For systems described by the Hamiltonian of Eq.~\eqref{GeneralHam}, we can directly evaluate the function $F_0(\mathbf{k})$ reported in Eq.~\eqref{F0} in both one and two dimensions. In particular, by inserting the components of $\mathbf{d_1}$ for the 1D case and of $\mathbf{d_2}$ for the 2D case, one explicitly obtains
\begin{equation}
    F_0^{(1D)}(\mathbf{k}) = k_x^2 (\Delta M)^2, 
    \qquad 
    F_0^{(2D)}(\mathbf{k}) = (k_x^2 + k_y^2) (\Delta M)^2,
\end{equation}
with $\Delta M \equiv m_A - m_B$. Putting all together, for $d-$dimensions} we get
\begin{equation}
    \Delta E_{d D}=\frac{S_{d-1}}{(2 \pi)^d}(\Delta M)^{2}\int_0^\Lambda \frac{k^{d+1}}{\left(k^2+m_{B}^{2}\right) \sqrt{k^2+m_{A}^{2}}} d k, \label{IntegraldD}
\end{equation}
with
\[S_{d-1} = \frac{2\pi^{\frac{d}{2}}}{\Gamma(\frac{d}{2})}\]
and $\Gamma(x)$ is the Gamma function. In the following section we will discuss the non-analytic behavior of this $d-$dimensional integral.\\ As we reduce our study to the region around $k = 0$, \textcolor{black}{focusing only on the singular part of Eq. \eqref{IntegraldD}}, we can simplify the integral by considering the limit $k << |m_A|$ as a valid approximation. The stored energy then becomes
\begin{equation}
    \Delta E_{dD}\simeq\frac{S_{d-1}}{(2 \pi)^d}\frac{(\Delta M)^{2}}{|m_A|}\int_0^\Lambda \frac{k^{d+1}}{k^2+m_{B}^{2}} d k.
\end{equation}
Since the integrand is a rational function, there exist two functions $Q(k)$ and $R(k)$ such that
\begin{equation}
        \int \frac{N(k)}{D(k)} d k=\int Q(k) d k+\int \frac{R(k)}{D(k)} d k
\end{equation}
with $Q(k)$ a polynomial in $k$ and $R(k)$ having a degree less than that of $D(k)$, so $R(k)$ must be either constant or linear in $k$. We now evaluate the integral for general values of $d$, and distinguish two cases based on its parity.
\subsection{Odd case: $d = 2n+1$}
By specifying the polynomial division in this case, we can observe that the integrand can be decomposed as
\begin{equation}
    \frac{k^{d+1}}{k^2+m_{B}^{2}} = \frac{k^{2 n+2}}{k^{2}+m_{B}^{2}}=\sum_{j=0}^{n}\left(-m_{B}^{2}\right)^{j} ~k^{2 n-2 j}+\frac{\left(-m_{B}^{2}\right)^{n+1}}{k^{2}+m_{B}^{2}}.
\end{equation}
Here, $Q(k) \equiv \sum_{j=0}^{n}\left(-m_{B}^{2}\right)^{j} k^{2 n-2 j}$ remains regular, since its $n$-th derivative does not develop any singularities as $m_B$ approaches zero. To investigate singularities, we then have to study the rational term $R/D(k)$ with $R \equiv \left(-m_{B}^{2}\right)^{n+1}$ constant in $k$. Integrals of this form can be solved in terms of an arctangent function; in particular
\begin{equation}
    \int_0^\Lambda \frac{R}{k^2 + m_B^2}dk = \frac{R}{m_B}\arctan\left(\frac{\Lambda}{m_B}\right).
\end{equation}
Substituting the form of $R$ and recalling that $2n+1 = d$, we have
\begin{equation}
    \frac{R}{m_B}\arctan\left(\frac{\Lambda}{m_B}\right) = (-1)^{\frac{d+1}{2}} m_B^d \arctan\left(\frac{\Lambda}{m_B}\right)
\end{equation}
and as $m_B$ approaches zero
\begin{equation}
    \arctan\left(\frac{\Lambda}{m_B}\right)\xrightarrow[m_{B} \rightarrow 0]{ } \frac{\pi}{2} \operatorname{sign}\left(m_{B}\right). \label{LimitArcTan}
\end{equation}
So, for odd values of $d$, the stored energy scales as $|m_B|^d$ and this leads to a jump discontinuity in its $d-$th derivative at $m_B = 0$. To evaluate the magnitude of this jump, let's take the complete form of the stored energy including its prefactor
\begin{equation}
    \Delta E_{dD}^{(odd)} \xrightarrow[m_{B} \rightarrow 0] {} \frac{S_{d-1}}{(2 \pi)^{d}} (\Delta M)^{2}(-1)^{\frac{d+1}{2}} \frac{\pi}{2} \frac{\left|m_{B}\right|^{d}}{\left|m_{A}\right|}.
\end{equation}
Let's consider a quench between an initial value $m_A$ and a final value $m_B = m_A + \delta$, with $\delta$ being a positive increment. By applying the chain rule for the $d$-th derivative
\begin{equation}
    \frac{d^{d}}{d m_{A}^{d}} \frac{\left|m_{B}\right|^{d}}{\left|m_{A}\right|}=\frac{d m_{B}^{d}}{d m_{A}^{d}} \frac{d}{d m_{B}^{d}} \frac{\left|m_{B}\right|^{d}}{\left|m_{A}\right|}=\frac{1}{\left|m_{A}\right|} d!\operatorname{sign}\left(m_{B}^{d}\right)
\end{equation}
and since $d$ is odd, then $\operatorname{sign}\left(m_{B}^{d}\right) = \operatorname{sign}\left(m_{B}\right) = \operatorname{sign}\left(m_{A} + \delta\right)$. The $d-$th derivative of the stored energy with respect to $m_A$ then becomes
\begin{equation}
    \frac{d^{d}}{d m_{A}^{d}} \Delta E_{dD}^{(odd)}=\frac{S_{d-1}}{(2 \pi)^{d}} ~\delta^{2}~(-1)^{\frac{d+1}{2}} \frac{\pi}{2} \frac{d!}{\left|m_{A}\right|} \operatorname{sign}\left(m_{A} + \delta\right)
\end{equation}
which evaluated for $m_A \to -\delta$ ($m_B \to 0$) gives the magnitude of the jump
\begin{equation}
    \left|\frac{S_{d-1}}{(2 \pi)^{d}} ~ (-1)^{\frac{d+1}{2}} ~\pi~ d! ~\delta\right| \label{d_DIM_JUMP}.
\end{equation}
\subsection{Even case: $d = 2n$}
In this case, the integrand can be written as
\begin{equation}
 \frac{k^{d+1}}{k^2+m_{B}^{2}} = \frac{k^{2 n+1}}{k^{2}+m_{B}^{2}}=\sum_{j=0}^{n-1}\left(-m_{B}^{2}\right)^{j} ~k^{2 n-1-2 j}+\frac{\left(-m_{B}^{2}\right)^{n} ~ k}{k^{2}+m_{B}^{2}}.
\end{equation}
Also in this scenario $Q(k) \equiv \sum_{j=0}^{n-1}\left(-m_{B}^{2}\right)^{j} ~k^{2 n-1-2 j}$ is regular, so we focus on the second term where $R(k) \equiv \left(-m_{B}^{2}\right)^{n} ~ k$ is linear in $k$. Introducing $\mathcal{C} \equiv \left(-m_{B}^{2}\right)^{n}$ the integral is
\begin{equation}
    \int_{0}^{\Lambda} \frac{\mathcal{C} k}{k^{2}+m_{B}^{2}} d k=\frac{\mathcal{C}}{2} \ln \left(1+\frac{\Lambda^{2}}{m_{B}^{2}}\right). \label{Int_pari}
\end{equation}
From Eq.~\eqref{Int_pari}, we can observe that as $m_B$ approaches zero, the term that leads to a singularity is proportional to \[m_B^d ~ \ln({|m_B|}).\]
Considering the same quench applied in the odd case, so from $m_A$ to $m_A + \delta$, applying the general Leibniz rule
\begin{equation}
    \frac{\partial^{n}}{\partial m_{A}^{n}}\left[(m_{A} + \delta)^{d} \ln \left|m_{A} + \delta\right|\right]=\sum_{k=0}^{n}\binom{n}{k} \frac{\partial^{n-k}}{\partial m_{a}^{n-k}} (m_{A} + \delta)^{d} \frac{\partial^{k}}{\partial m_{a}^{k}} \ln \left|m_{A} + \delta\right|
\end{equation}
we obtain after a straightforward calculation
\begin{equation}
\begin{aligned}
    &\frac{d^{n}}{d m_{A}^{n}}\left[(m_{A} + \delta)^{d} \ln \left|m_{A} + \delta\right|\right]=\\=&(m_{A} + \delta)^{d-n}\left[\frac{d!}{(d-n)!} \ln \left|m_{A} + \delta\right|+d!\sum_{k=1}^{n}\binom{n}{k}(-1)^{k-1} \frac{(k-1)!}{(d-n+k)!}\right] \label{GLR}
\end{aligned}
\end{equation}
from which we can observe that if $n < d$, then $(m_A+\delta)^{d-n} ~ \ln{|m_A+\delta|} \xrightarrow[m_A \to -\delta]{}0$ and no singularities appear. However, if $n = d$, then Eq.\eqref{GLR} becomes
\begin{equation}
    \frac{d^{d}}{d m_{A}^{d}}\left[(m_{A} + \delta)^{d} \ln \left|m_{A} + \delta\right|\right] = d!\left(\ln{|m_A + \delta|} + \mathcal{H}_d\right) \label{GLRconNequalD}
\end{equation}
where \[\mathcal{H}_d = \sum_{k=1}^d \binom{d}{k} \frac{(-1)^{k-1}}{k}\] is the harmonic number. From Eq.~\eqref{GLRconNequalD} the singularity is evident, since $\ln{|m_A + \delta|}$ diverges as $m_A$ approaches $-\delta$. Therefore, in both scenarios the $d-$th derivative of the stored energy shows a singularity in the limit $m_A \to -\delta$, but its nature is different in the two cases: for odd values of $d$ the non-analyticity is given by a finite jump, while for even values of $d$ we observe a logarithmic divergence. The explicit calculation for $d=1,2$ are shown in Appendix I and II respectively. We now provide a one- and two-dimensional model examples in the full brillouin zone to support the equations we derived and to address the numerical aspects of these singularities.


\section{Lattice models}
In this section, we check our results on the basis of two lattice models, namely the quantum Ising chain in a transverse field and the Haldane model.
\subsection{Ising chain}
\begin{figure}[H]
    \centering
    \includegraphics[width=0.7\linewidth]{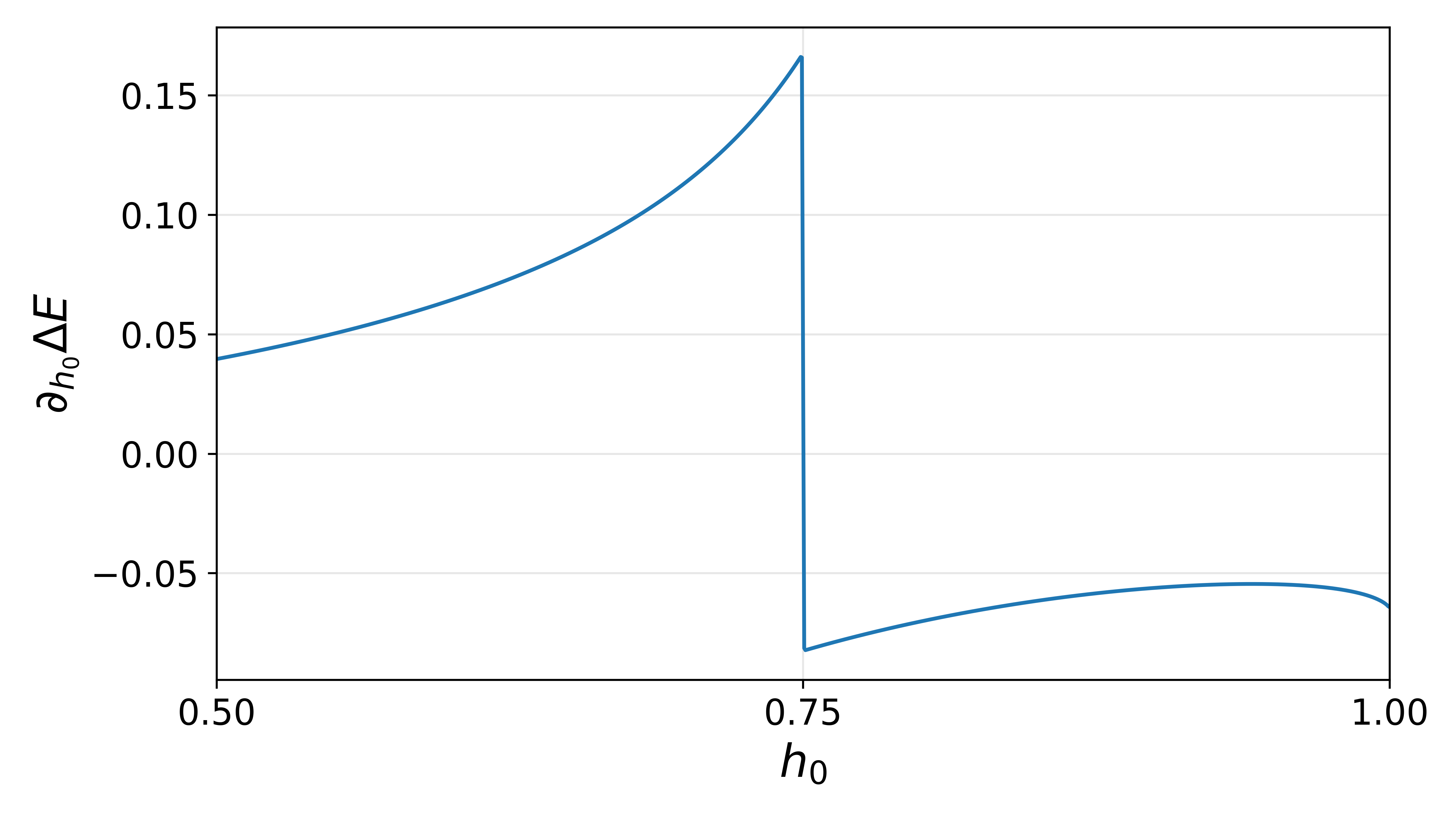}
    \caption{First derivative of the stored energy in the Ising chain with respect to the transverse field as a function of the initial magnetic field $h_0$.}
    \label{fig:ising}
\end{figure}

As stated above the quantum Ising chain in a transverse field~\cite{Pfeuty70} is the paradigmatic model for the study of quantum phase transitions.
Its Hamiltonian reads as
\begin{equation}
H_I=-\sum_{j=1}^{N} \left[\sigma^x_{j+1}\sigma^x_{j}+h\sigma^z_{j}\right].
\end{equation}
Here, $j$ enumerates the lattice sites of a chain with a total of $N$ sites, $\sigma_j^{x/z}$ are the Pauli matrices on the site $j$, and $h$ is the external field. Periodic boundary conditions are imposed. The ground state diagram has two phases: a paramagnetic one, for applied transverse fields $|h|>1$ and a ferromagnetic one for $|h|<1$. The transition between the two phases happens through a second order quantum phase transition. A remarkable feature of the quantum Ising chain in a transverse field is that it is exactly solvable by means of a Wigner-Jordan transformation to fermions (followed by a Fourier expansion to exploit translational invariance and a Bogoliubov transformation for getting rid of the superconducting correlations). For the details we refer for instance to Ref.~\cite{Franchini16}. What we consider here is the quantum Ising chain in a transverse field as a quantum battery. As a charging protocol we adopt, as usual, a double sudden quantum quench where a parameter is varied at $t=0$, and at $t=\tau$ it is set back to its initial value. In particular, here we switch between the values $h_0$ and $h_0 + h_1$ of the applied magnetic field. For the energy stored $\Delta E^{Ising}$ one has, at long times,
\begin{equation}
    \Delta E^{Ising} = h_1^2\sum_{k \in BZ}\frac{\sin^2(k)}{2\epsilon^{Ising}(k)\left(\omega^{Ising}(k)\right)^2}.
\end{equation}
Here the dispersion relations are given by
\begin{eqnarray}
\epsilon^{Ising}(k)&=&\sqrt{(h_0-\cos(k))^2+\sin^2(k)},\\
\omega^{Ising}(k)&=&\sqrt{(h_0+h_1-\cos(k))^2+\sin^2(k)}.
\end{eqnarray}
From the equation above one can check that the first derivative of the energy stored is characterized by a finite jump in correspondence of the quantum phase transition $h_0+h_1=1$. The magnitude of the jump is in accordance with the low energy theory developed by concentrating on the physics at the cone described in the previous section. This fact is reported in Fig. \ref{fig:ising}. There, the parameter used is $h_1 = 0.25$, so that the reported jump of magnitude $h_1$ at $h_0 = 0.75$ is indeed consistent with the previous discussion (for more details, we refer to Eq. \eqref{1D_Jump_Appendix} in Appendix I).

\subsection{Haldane model}
\begin{figure}[H]
    \centering
    \includegraphics[width=0.5\linewidth]{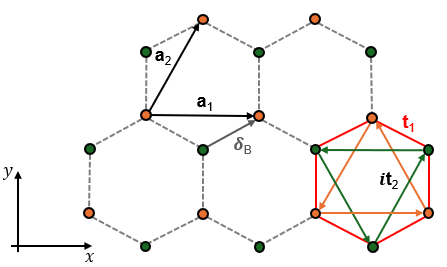}
    \caption{Honeycomb lattice of the Haldane model.}
    \label{Fig_Haldane}
\end{figure}

\begin{figure}[H]
    \centering
    \includegraphics[width=0.9\linewidth]{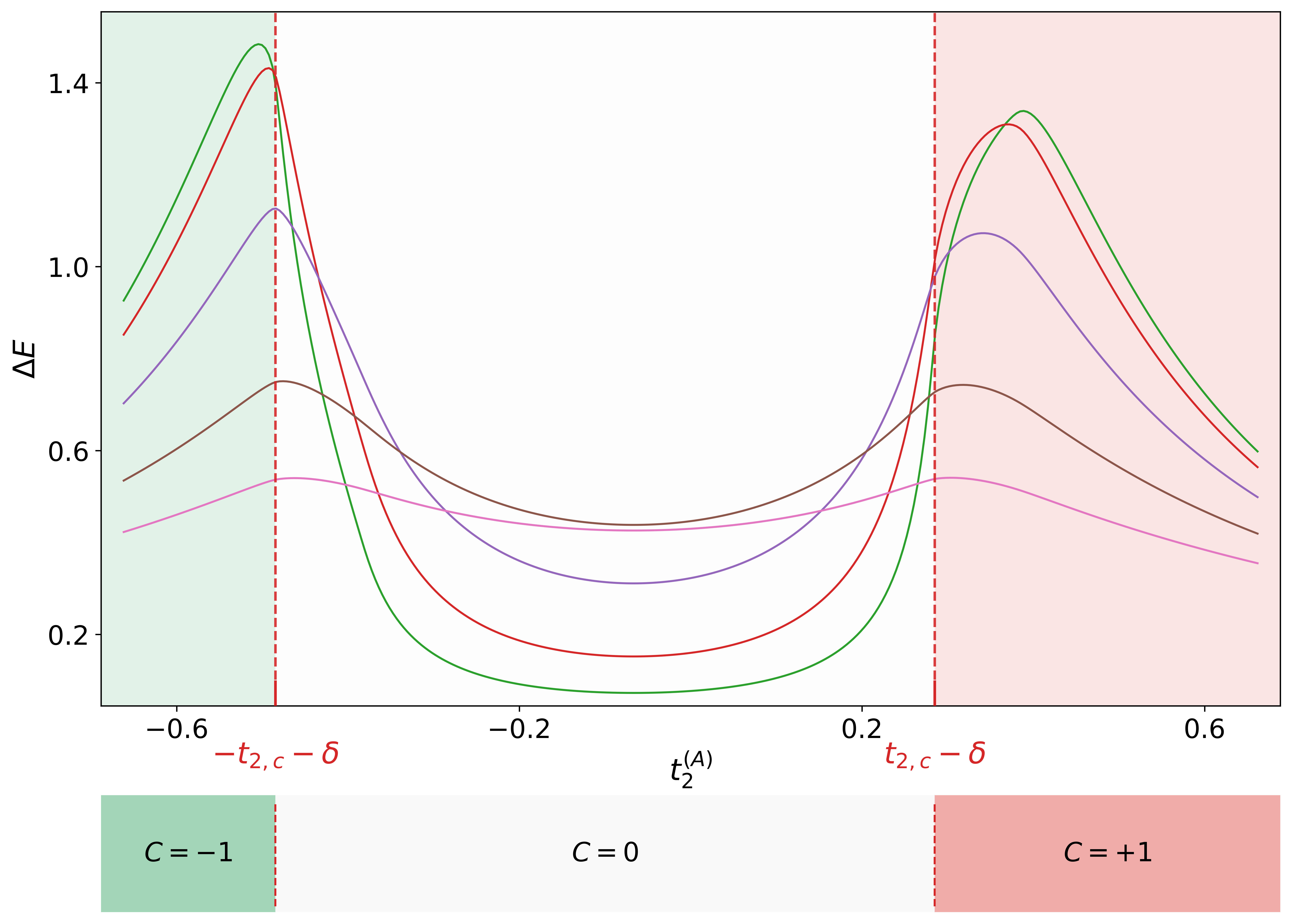}
    \caption{\textcolor{black}{Energy stored in the Haldane model as function of $t_2^{(A)}$ for $t_1 = 0.5$ (green), $t_1 = 0.8$ (red), $t_1 = 1.5$ (purple), $t_1 = 3$ (brown) and $t_1 = 5$ (pink) with fixed $\delta = 0.1$.} The red dotted lines represent the critical $t_2$ values for the evolution Hamiltonian, while the different colors indicate the phase of the model according to the Chern number's value, as represented under the plot: topological phase with $C = -1$ (green region), topological phase with $C = 1$ (red region) and trivial phase with $C = 0$ (white region)}
    \label{Haldane_quench_t1_less_1}
\end{figure}

An important two-dimensional model we can study in a similar way is the one introduced by F.~D.~M.~Haldane in 1988~\cite{Haldane88}. While this model is certainly paradigmatic in the topological material community, it might be not extremely common in terms of its energetic aspects, so that we briefly review its properties before addressing the original results.\\
The Haldane model is a tight-binding model defined on a honeycomb lattice (which can be viewed as a triangular Bravais lattice with two atoms, type $A$ and type $B$, per unit cell) that demonstrates how a system can exhibit a quantized Hall conductance without an external magnetic field. \textcolor{black}{The key idea is that a quantized Hall response does not require a uniform magnetic field or Landau levels: it instead requires a nontrivial distribution of Berry curvature in momentum space and a resulting nonzero Chern number, i.e. the topological invariant characterizing the band structure of two dimensional quantum Hall systems ~\cite{Hasan10}, for the occupied band. In the Haldane model this is achieved by engineering a local time-reversal-symmetry breaking on the lattice while keeping the net magnetic flux through each unit cell equal to zero. This produces an effective magnetic field in momentum space, concentrated around the gapped Dirac points, whose integral over the Brillouin zone yields the quantized Hall conductance \cite{Wang2008, Thouless1982, Minkov16, Xiao2010}.} A paradigmatic system that presents a honeycomb lattice is graphene~\cite{CastroNeto09}: here the carbon atoms interact through a nearest-neighbor hopping term with amplitude $t_1$, and the corresponding Hamiltonian is characterized by gapless Dirac cones and massless quasiparticles. \textcolor{black}{As stated above,} the Haldane model extends this Hamiltonian by introducing both a next-nearest-neighbor hopping term with amplitude $t_2 e^{i\phi}$ between sites on the same sublattice, where the complex phase $\phi$ associated with this hopping breaks time-reversal symmetry and introduces a mass term at the Dirac points (from now on we will set $\phi = \frac{\pi}{2}$), and a staggered on-site potential $m$, which takes opposite signs on the two sublattices and breaks inversion symmetry. These perturbations gap out the Dirac cones and give rise to a topologically nontrivial insulating phase characterized by a nonzero Chern number. We pick as primitive vectors $\mathbf{a}_1 = a(1,0)$ and $\mathbf{a}_2 = a(\frac{1}{2}, \frac{\sqrt{3}}{2})$, with $a$ being the Bravais lattice spacing. By placing the origin of the unit cell on the $A$ sites, the basis vectors are $\boldsymbol{\delta}_A = (0,0)$ and $\boldsymbol{\delta}_B = a(\frac{1}{2}, \frac{1}{2\sqrt{3}})$. In Fig. \ref{Fig_Haldane} a sketch of the lattice is shown: for clarity, we have represented the lattice vectors and the hoppings in two different hexagons.
In $\mathbf{k}-$space, the Hamiltonian of the Haldane model, for our choice of $\phi$, assumes the usual form $H(\mathbf{k}) = \mathbf{d}(\mathbf{k}) \cdot \boldsymbol{\sigma}$, where 
\begin{equation}
    \begin{aligned}
        &d_1(\mathbf{k}) = t_1 \left[\cos(\mathbf{k} \cdot \boldsymbol{\delta}_B) + \cos\left(\mathbf{k} \cdot (\boldsymbol{\delta}_B - \mathbf{a}_1)\right) + \cos\left(\mathbf{k} \cdot (\boldsymbol{\delta}_B - \mathbf{a}_2)\right)\right]\\
        &d_2(\mathbf{k}) = -t_1 \left[\sin(\mathbf{k} \cdot \boldsymbol{\delta}_B) + \sin\left(\mathbf{k} \cdot (\boldsymbol{\delta}_B - \mathbf{a}_1)\right) + \sin\left(\mathbf{k} \cdot (\boldsymbol{\delta}_B - \mathbf{a}_2)\right)\right] \\
        &d_3(\mathbf{k}) = m + 2t_2\left[-\sin(\mathbf{k} \cdot \mathbf{a}_1) + \sin\left(\mathbf{k} \cdot (\mathbf{a}_1 - \mathbf{a}_2)\right) + \sin\left(\mathbf{k} \cdot \mathbf{a}_2\right)\right].
    \end{aligned}
\end{equation}
Note that a ($k$-dependent) energy shift has been neglected as it does not alter the results.
The Dirac points are obtained from the condition that the off-diagonal part of the Bloch Hamiltonian vanishes, so we impose
\begin{equation}
d_1(\mathbf{k}) = 0,\qquad d_2(\mathbf{k}) = 0 .
\end{equation}
Equivalently, one can write
\begin{equation}
d_1(\mathbf{k}) + i d_2(\mathbf{k}) = t_1\sum_{j=1}^3 e^{\textcolor{black}{-}i\mathbf{k}\cdot\boldsymbol{\delta}_j}=0,
\end{equation}
where we used the three nearest-neighbour vectors
\begin{equation}
\boldsymbol{\delta}_1=\boldsymbol{\delta}_B,\qquad
\boldsymbol{\delta}_2=\boldsymbol{\delta}_B-\mathbf{a}_1,\qquad
\boldsymbol{\delta}_3=\boldsymbol{\delta}_B-\mathbf{a}_2 .
\end{equation}
A vanishing sum of these three complex exponentials requires that their phases differ by \(\pm 2\pi/3\); this means that we have the following linear constraints
\begin{equation}
\mathbf{k}\cdot(\boldsymbol{\delta}_2-\boldsymbol{\delta}_1)=\frac{2\pi}{3},\qquad
\mathbf{k}\cdot(\boldsymbol{\delta}_3-\boldsymbol{\delta}_1)=-\frac{2\pi}{3},
\end{equation}
and since \(\boldsymbol{\delta}_2-\boldsymbol{\delta}_1=-\mathbf{a}_1\) and \(\boldsymbol{\delta}_3-\boldsymbol{\delta}_1=-\mathbf{a}_2\) these become
\begin{equation}
\mathbf{k}\cdot\mathbf{a}_1=-\frac{2\pi}{3},\qquad
\mathbf{k}\cdot\mathbf{a}_2=+\frac{2\pi}{3}.
\end{equation}
Writing \(\mathbf{k}=\alpha\mathbf{b}_1+\beta\mathbf{b}_2\) in the reciprocal-lattice basis (with \(\mathbf{b}_i\cdot\mathbf{a}_j=2\pi\delta_{ij}\)) yields $\alpha=-\frac{1}{3}$ and $\beta=-\frac{1}{3}$. Therefore, one Dirac point is
\begin{equation}
\mathbf{K}=\frac{-\mathbf{b}_1+\mathbf{b}_2}{3},
\end{equation}
and the other one is \(\mathbf{K}' =-\mathbf{K}\). Considering the primitive vectors we chose, the corresponding reciprocal vectors are
\begin{equation}
\mathbf{b}_1=\frac{2\pi}{a}\Big(1,-\tfrac{1}{\sqrt{3}}\Big),\qquad
\mathbf{b}_2=\frac{2\pi}{a}\Big(0,\tfrac{2}{\sqrt{3}}\Big)
\end{equation}
and the Dirac points assume the form
\begin{equation}
\mathbf{K}=\frac{\mathbf{b}_2-\mathbf{b}_1}{3}=\frac{2\pi}{3a}\,(-1,\sqrt{3}),\qquad
\mathbf{K}'=-\mathbf{K}=\frac{2\pi}{3a}\,(1,-\sqrt{3}).
\end{equation}
If we now evaluate \(d_3(\mathbf{k})\) at \(\mathbf{K},\mathbf{K}'\) we obtain the two Dirac masses
\begin{equation}
m_{\mathbf{K}} \;=\; m - 3\sqrt{3}\,t_2,\qquad
m_{\mathbf{K}'} \;=\; m + 3\sqrt{3}\,t_2. \label{masses}
\end{equation}
These masses arise from the complex next-nearest-neighbor hopping, which breaks time-reversal symmetry and generates opposite signs at the two Dirac points. In particular, Eq.~\eqref{masses} follows from our choice of \(\phi = \frac{\pi}{2}\); more generally, any phase \(\phi\) different from \(0\) and \(\pi\) leads to masses of opposite sign and is responsible for the emergence of the Chern-insulating phase. Finally, it is worth noticing that by expanding the Hamiltonian around the Dirac points, one obtains the low-energy effective theory of the Haldane model, which reduces to a massless Dirac Hamiltonian. Concerning the QPTs, they occur when either \(m_{\mathbf{K}}\) or \(m_{\mathbf{K}'}\) changes sign, i.e., when one of the two masses becomes zero. This condition yields the critical value of \(t_2\), namely
\begin{equation}
    t_{2,c} = \pm \frac{m}{3\sqrt{3}} \, .
\end{equation}
For simplicity, we fix $m = 1$ so that we expect a QPT at $t_{2,c} = \pm \frac{1}{3\sqrt{3}} \approx \textcolor{black}{\pm} 0.1925$. A compact and practical expression for the Chern number is
\begin{equation}
C=\frac{\operatorname{sgn}(m_{\mathbf{K}'})-\operatorname{sgn}(m_{\mathbf{K}})}{2}\,.
\end{equation}
Therefore, we can distinguish three different phases according to the value of $C$:
\begin{itemize}
  \item \(t_2 < -t_{2,c}\): topological phase with \(C=-1\).
  \item \(-t_{2,c} < t_2 < t_{2,c}\): trivial phase with \(C=0\).
  \item \(t_2 > t_{2,c}\): topological phase with \(C=+1\).
\end{itemize}
We now come to the Haldane model as a quantum battery. In analogy with the previous models we perform a quench of the $t_2$ parameter of the form $t_2^{(A)} \to t_2^{(A)} + \delta \to t_2^{(A)}$, with fixed $\delta = 0.1$, to study what happens at the critical points $\pm t_{2,c} - \delta$ of the evolution Hamiltonian. Since the model is more complex than the Dirac one, our analysis takes into account different values of the nearest-neighbor hopping $t_1$. The plots shown in the following figures highlights interesting results: first, as we observe in Fig. \ref{Haldane_quench_t1_less_1}, for $t_1 \leq m$ the energy peaks always appear when the model is either in its topological phase or on the critical line, highlighting the distinct behavior associated with each phase of the system. As $t_1$ increases, we observe in \textcolor{black}{the same plot} that the curves gradually flatten, while the energy decreases and approaches zero. This latter trend can be qualitatively explained as follows: for large $t_1$, the Haldane Hamiltonian reduces to the massless Dirac Hamiltonian describing graphene, since the dynamics are dominated by the nearest-neighbor hopping term. In this regime, quenching $t_2$, which only enters the mass term, has a negligible effect on the Hamiltonian, and thus the system stores almost no energy, as if no quench had occurred.

\begin{figure}[H]
    \centering
    \includegraphics[width=0.9\linewidth]{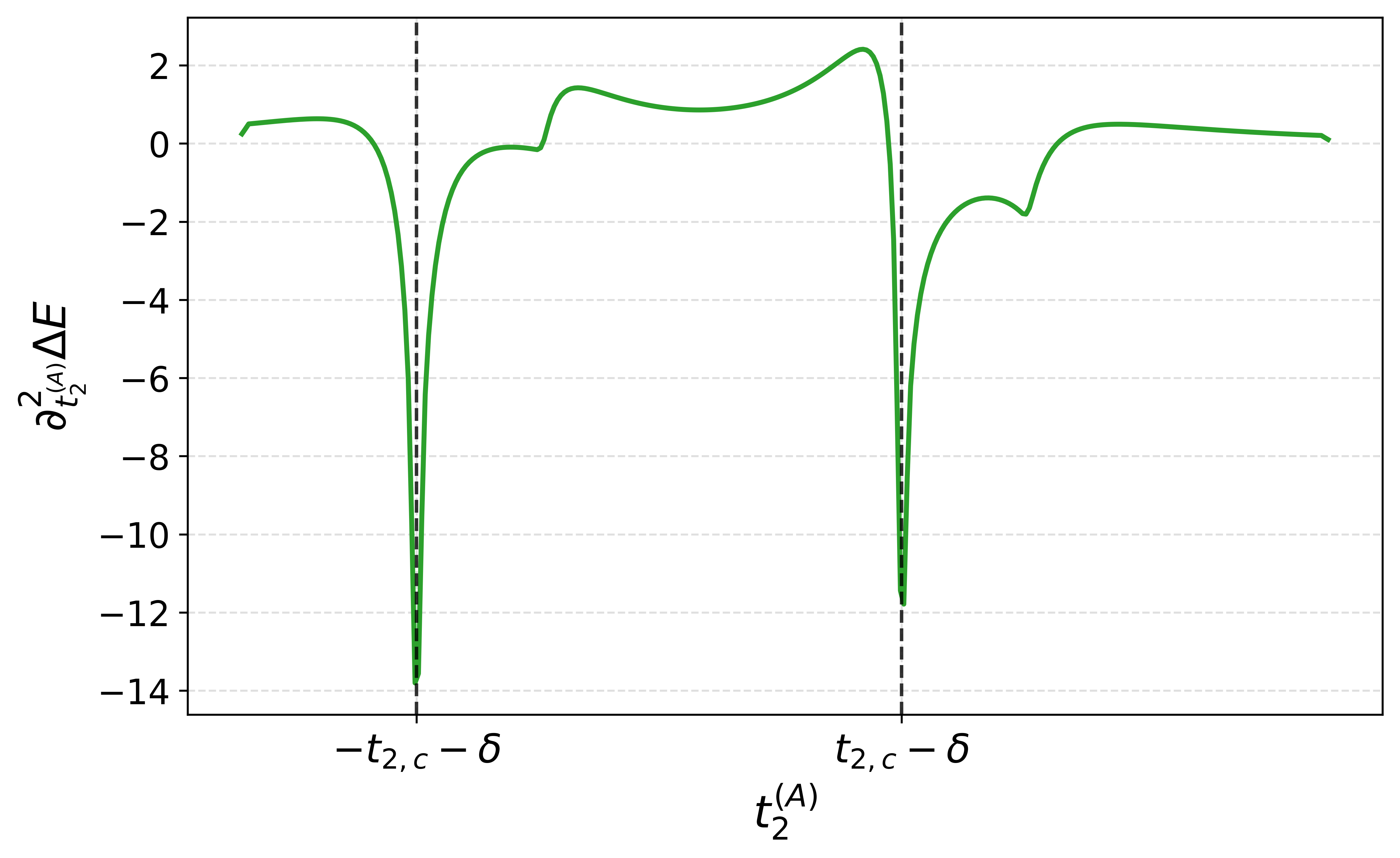}
    \caption{Second derivative of the energy stored in the Haldane model as a function of $t_2^{(A)}$.}
    \label{fig:hal}
\end{figure}

Let us now come to the comparison between the energy stored in the Haldane model quantum battery and the discussion of the previous section. The second derivative of the energy stored with respect to $t_2^{(A)}$ is shown in Fig. \ref{fig:hal}. There, the values of the parameters are $t_1 = m = 1$ and $\delta = 0.1$, so that the logaritmic divergencies in correspondence of the quantum phase transitions are evident (for more details, we refer to Eq. \eqref{2D_Log_Div} in Appendix II).


\section{Conclusions}
We have considered the energy stored in many-body free fermion quantum batteries. Such batteries are interesting because they represent a large class of models, that range from Wigner-Jordan integrable spin chains, to lattice models with genuinely fermionic degrees of freedom. The emphasis is on the role on quantum phase transitions on this quantity. Indeed, it was previously shown that quantum phase transitions can help increasing or stabilizing the energy stored, depending on the model. Moreover, the presence of quantum phase transitions in the quantum system used as a quantum battery also leads to non-analiticities in the energy stored. In this work we have analyzed the universal features related to quantum phase transitions in free fermionic quantum batteries. To do so, we have isolated the contribution of 'the Dirac cone'. This allowed us to prove that in odd dimensions the presence of quantum phase transitions is accompained by a finite jump in the $d-$th derivative of the energy stored. In even dimensions, on the other hand, the non-analytical behavior is a logaritmic divergence of the $d-$th derivative of the energy stored. \textcolor{black}{These non-analyticities originate from the enhanced contribution of low-energy modes when the post-quench Hamiltonian approaches a gap closing, which makes the stored energy extremely sensitive to small changes of the control parameters. This critical enhancement of susceptibility shows a trade-off between stability and controllability. Indeed, on the one hand, operating close to a critical point reduces the stability of the battery, as small imperfections or noise in the protocol can produce large variations in the stored energy; on the other hand, the same sensitivity can be exploited to improve the control, since precise and intentional variations of the quenched parameter can induce strong energetic responses with a potentially positive impact of the permances of the device.} We have then tested our results with two low energy models: the quantum Ising chain in a transverse field as regards one dimension, and the Haldane model for a Chern insulator as a representative of two dimensional systems. The results confirm what derived in the low energy approximation. Moreover, interstingly, the energy stored in the Haldane model intriguingly follows the topological phase diagram of the system, opening interesting perspectives for an energy related detection of topological phases possibly applicable to some specific models.

\appendix
\section{Appendix I: 1-D Dirac model}
\begin{figure}[H]
    \centering
    \includegraphics[width=0.7\linewidth]{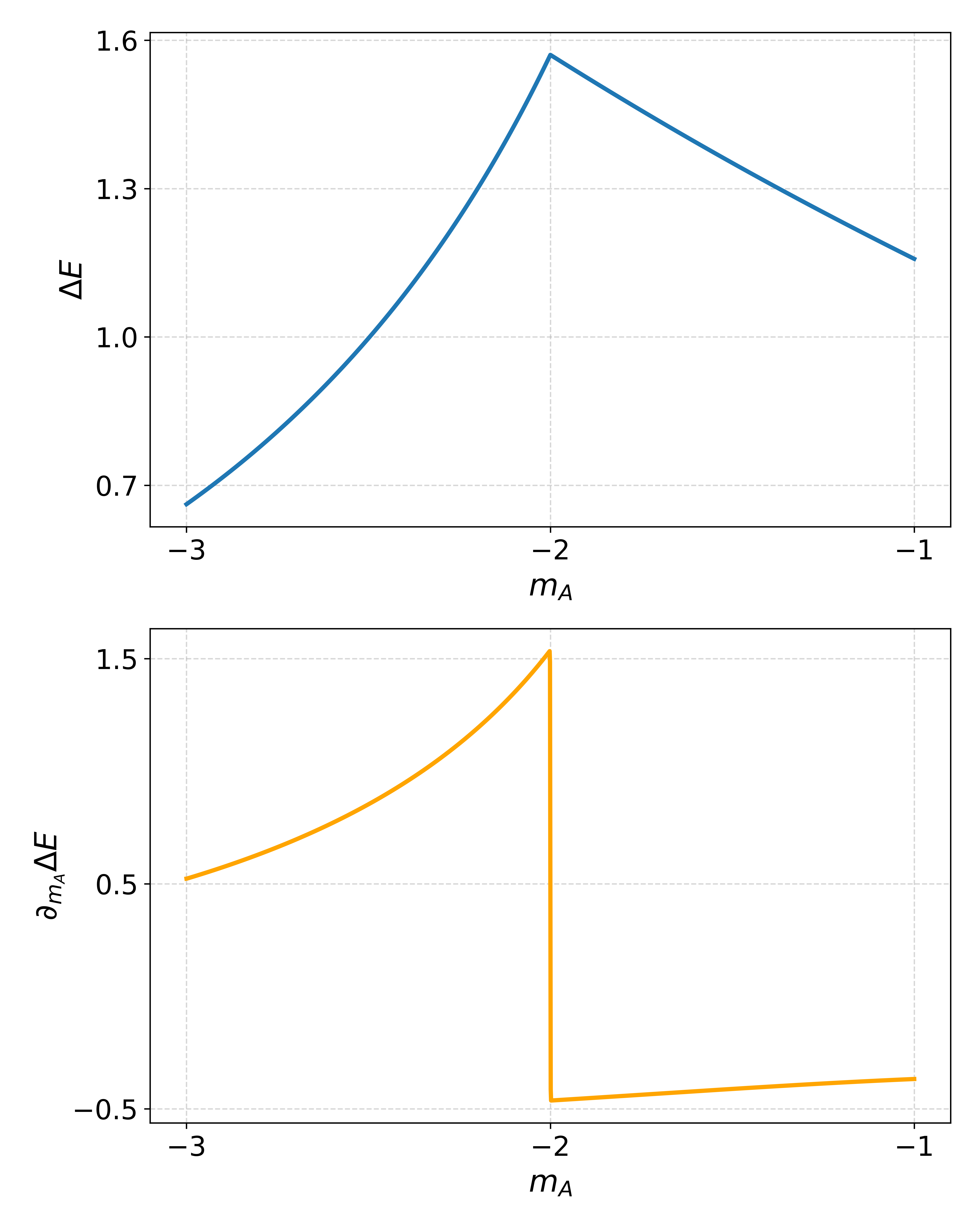}
    \caption{\textcolor{black}{Energy stored in the 1-D Dirac model as function of $m_A$ for $\delta = 2$ (blue curve) and jump in its first derivative with respect to $m_A$ as function of $m_A$ (orange curve).}}
    \label{1D_Energy_Derivative}
\end{figure}

In one dimension the stored energy around $k = 0$ is given by the integral
\begin{equation}
    \Delta E_{1D}=\frac{(\Delta M)^{2}}{\pi}\int_0^\Lambda \frac{k^2}{\left(k^2+m_{B}^{2}\right) \sqrt{k^2+m_{A}^{2}}} d k. \label{Integral1D}
\end{equation}
In the limit $k << |m_A|$ the integral can be solved as follows
\begin{equation}
    \Delta E_{1D} \simeq \frac{(\Delta M)^{2}}{\pi |m_A|}\int_0^\Lambda \frac{k^2}{k^2+m_{B}^{2}} d k = \frac{(\Delta M)^{2}}{\pi |m_A|}  \left[\Lambda - m_B ~ \arctan{\left(\frac{\Lambda}{m_B}\right)}\right].
\end{equation}
By using Eq.~\eqref{LimitArcTan} we obtain
\begin{equation}
    \Delta E_{1D} \xrightarrow[m_B \to 0]{} \frac{(\Delta M)^{2}}{\pi |m_A|}\Lambda - \frac{(\Delta M)^{2}}{2|m_A|}|m_B|.
\end{equation}
By substituting $m_B = m_A + \delta$, the first derivative of the stored energy with respect to $m_A$ gives
\begin{equation}
    \frac{\partial \Delta E_{1D}}{\partial m_A} = (\text{regular terms}) - \frac{\delta^2}{2|m_A|}\operatorname{sign}(m_A + \delta).
\end{equation}
As predicted, in the limit $m_A \to -\delta$ the discontinuity arises in the first derivative (being $d = 1$) and the jump is
\begin{equation}
    \frac{\delta^2}{|m_A|} = \frac{\delta^2}{\delta} = \delta.
\end{equation}
This result is in agreement with the general formula shown in Eq.\eqref{d_DIM_JUMP}, since for $d = 1$ it becomes
\begin{equation}
    \left|-\frac{S_{0}}{2 \pi} ~\pi ~\delta\right| = \delta \label{1D_Jump_Appendix}
\end{equation}
where we used the fact that
\begin{equation}
    S_0 =\frac{2\sqrt{\pi}}{\Gamma(\frac{1}{2})} = \frac{2\sqrt{\pi}}{\sqrt{\pi}} = 2.
\end{equation}
From the numerical analysis, we can observe \textcolor{black}{in Fig. \ref{1D_Energy_Derivative}} that, for fixed $\delta = 2$, the stored energy shows a peak at $m_A = -\delta$, in accordance with our previous studies~\cite{Grazi24, Grazi25, Grazi25b}, while its first derivative shows the jump discontinuity we predicted in the previous section, with the magnitude of such jump being exactly $\delta$.
\section{Appendix II: 2-D Dirac model}
\begin{figure}[H]
    \centering
    \includegraphics[width=0.8\linewidth]{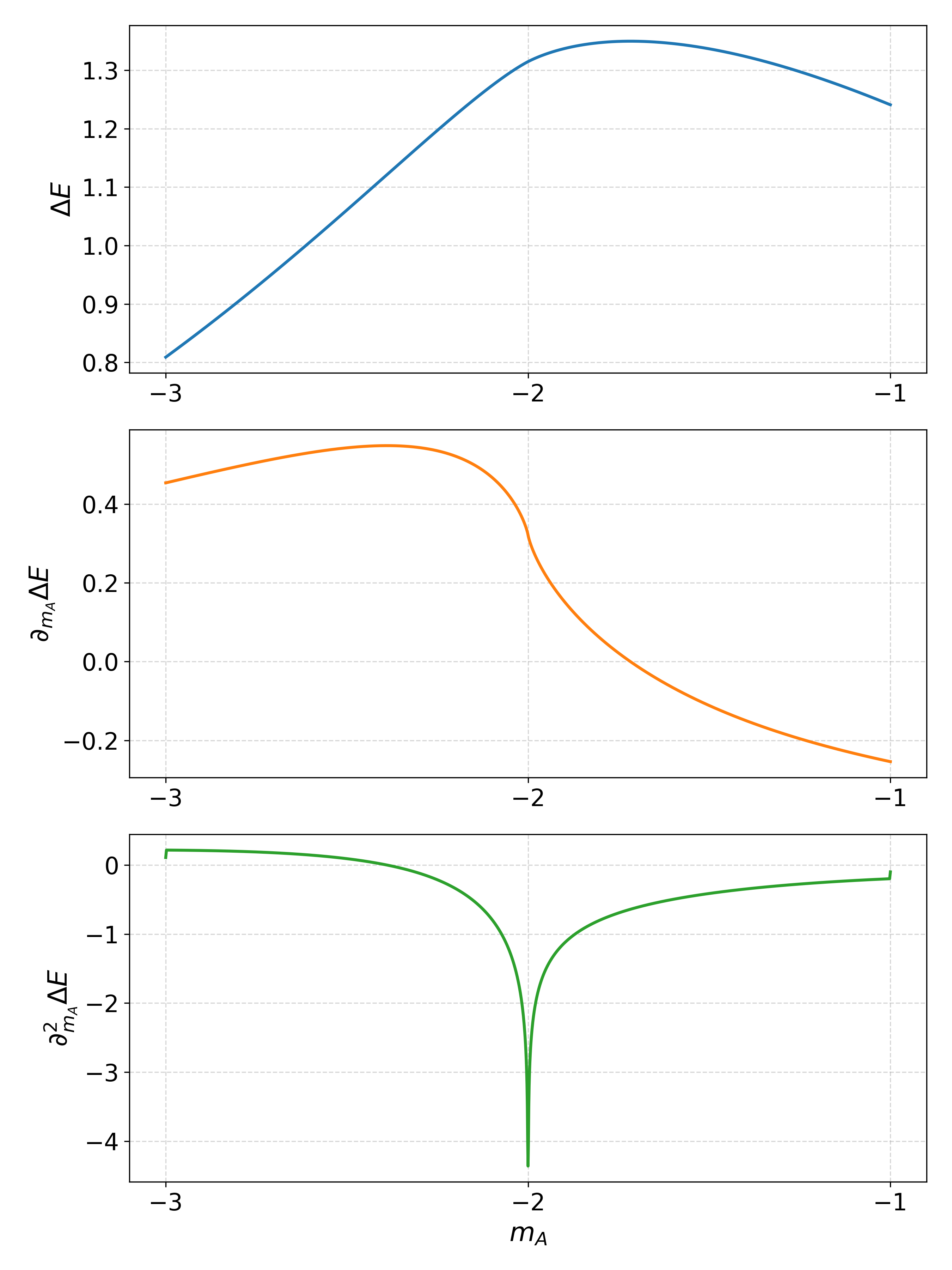}
    \caption{Stored energy in the 2-D Dirac model (blue), its first (orange) and second derivatives (green) as functions of $m_A$.}
    \label{Energy_2D_Plots}
\end{figure}

Here, we aim at studying the energy stored in the Hamiltonian reported in Eq. ~\eqref{GeneralHam} by solving the integral of Eq.~\eqref{IntegraldD} for $d=2$
\begin{equation}
    \Delta E_{2D}=\frac{(\Delta M)^{2}}{2 \pi |m_A|}\int_0^\Lambda \frac{k^3}{k^2+m_{B}^{2}} d k.
\end{equation}
where we considered $k << |m_A|$ as in the previous case. Once we solve the integral
\begin{equation}
    \int_0^\Lambda \frac{k^3}{k^2+m_{B}^{2}} d k = \Lambda^2 - m_B^2 \ln{\left(\frac{\Lambda^2 + m_B^2}{m_B^2}\right)}
\end{equation}
we consider the limit of $m_B$ approaching zero. Here
\begin{equation}
    \ln{\left(\frac{\Lambda^2 + m_B^2}{m_B^2}\right)} \approx \ln{\left(\frac{\Lambda^2}{m_B^2}\right)} = 2 \ln{(\Lambda)} - 2 \ln{(|m_B|)}
\end{equation}
and we obtain
\begin{equation}
    \Delta E_{2D} \xrightarrow[m_B \to 0]{} \frac{(\Delta M)^{2}}{4 \pi |m_A|}\Lambda^2 - \frac{(\Delta M)^{2}}{2 \pi |m_A|} m_B^2 \ln{(\Lambda)} + \frac{(\Delta M)^{2}}{2 \pi |m_A|}m_B^2 \ln{(|m_B|)}
\end{equation}
whose derivative with respect to $m_A$, after substituting $m_B = m_A + \delta$, is
\begin{equation}
    \frac{\partial \Delta E_{2D}}{\partial m_A} = \text{(regular terms)} + \frac{\delta^{2}}{2 \pi |m_A|}\left[\frac{m_A(m_A+\delta) + (m_A-\delta)(m_A+\delta)\ln{|m_A+\delta|}}{m_A}\right].
\end{equation}
Unlike in the one-dimensional case, here the first derivative does not exhibit singularities as $m_A \to -\delta$, since
\[
(m_A+\delta)\ln{|m_A+\delta|} \xrightarrow[m_A \to -\delta]{}0
\]
by L'H\^opital's theorem. However, if we derive again, we obtain
\begin{equation}
    \frac{\partial^2 \Delta E_{2D}}{\partial m_A^2} = \text{(regular terms)} + \frac{\delta^{2}}{2 \pi |m_A|}\left[2\ln{|m_A+\delta| + 3}\right]. \label{2D_Log_Div}
\end{equation}
and since $\ln{|m_A+\delta|}$ diverges as $m_A$ approaches $-\delta$, we observe the logarithmic divergence we predicted in Section \ref{Dirac}. The numerical plots in Fig. \ref{Energy_2D_Plots} show the stored energy (blue curve), its first derivative (orange curve) and its second derivative (green curve) with respect to $m_A$ for fixed $\delta = 2$. The derivative plots are consistent with our theoretical analysis, and this behavior is reflected in the energy plot, where in this case no peak appears at $m_A = -\delta$, unlike in the one-dimensional scenario. In contrast, the maximum energy is reached after the QPT, highlighting the advantage of undergoing a phase transition to store more energy.

\subsection*{Fundings}
N.T.Z. acknowledges funding through the “Non-reciprocal supercurrent and topological transitions in hybrid Nb- InSb nanoflags” project (Prot. 2022PH852L) in the framework of PRIN 2022 initiative of the Italian Ministry of University (MUR) for the National Research Program (PNR). This project has been funded within the programme “PNRR Missione
4—Componente 2—Investimento 1.1 Fondo per il Programma Nazionale di Ricerca e Progetti
di Rilevante Interesse Nazionale (PRIN)”. D.F. acknowledges funding from the European Union-
NextGenerationEU through the the “Solid State Quantum Batteries: Characterization and Optimization” (SoS-QuBa) project (Prot. 2022XK5CPX), in the framework of the PRIN 2022 initiative of the Italian Ministry of University
(MUR) for the National Research Program (PNR). This project has been funded within the program
“PNRR Missione 4—Componente 2—Investimento 1.1 Fondo per il Programma Nazionale di Ricerca
e Progetti di Rilevante Interesse Nazionale (PRIN)”.

\bibliographystyle{unsrt}
\bibliography{Universal-bibliography}

\end{document}